\documentclass[doublecol,showpacs,aps,prl,floatfix,superscriptaddress]{epl2}
\usepackage{graphicx}
\usepackage{amsmath,amssymb,amsfonts,mathrsfs}
\usepackage{epic,eepic}
\usepackage{graphics}
\usepackage{epsfig}

\usepackage{subfigure}

\def\id{{1 \kern-.30em 1}}

\newcommand{\be}{\begin{equation}}
\newcommand{\ee}{\end{equation}}
\newcommand{\bea}{\begin{eqnarray}}
\newcommand{\eea}{\end{eqnarray}}

\usepackage{amsmath}
\usepackage{amsfonts}
\usepackage{amssymb}
\usepackage{bm}
\usepackage{color}
\usepackage{graphicx}
\usepackage{breqn}

\title{Rheological properties of soft-glassy flows from hydro-kinetic simulations}
\shorttitle{Soft-glassy flows from hydro-kinetic simulations}

\author{R. Benzi \inst{1}, M. Bernaschi\inst{2}, M. Sbragaglia\inst{1}, S. Succi\inst{2} }
\shortauthor{R. Benzi \etal}

\institute{
  \inst{1} Department of Physics and INFN, University of ``Tor Vergata'', Via della Ricerca Scientifica 1, 00133 Rome, Italy\\
  \inst{2} Istituto per le Applicazioni del Calcolo CNR, Via dei Taurini 19, 00185 Rome, Italy}


\abstract{Based on numerical simulations of a lattice kinetic model for soft-glassy materials, we characterize the global rheology of a dense emulsion-like system, under three representative load conditions: Couette flow, time-oscillating Strain and Kolmogorov flow.  It is found that in all cases the rheology is described by a Herschel-Bulkley (HB) relation, $\sigma = {\sigma}_{Y} + A S^{\beta}$,  with the yield stress ${\sigma}_{Y}$ largely independent of the loading scenario.  A proper rescaling of the HB parameters permits to describe heterogeneous flows with space-dependent stresses, based on the notion of cooperativity, as recently proposed to characterize the degree of non-locality of stress relaxation phenomena in soft-glassy materials. }

\begin{document}

\maketitle

\section{Introduction}
Soft-Glassy (SG) materials represent one of the major open problems in modern thermodynamics and non equilibrium statistical mechanics \cite{Larson,Coussot}.
Despite their widely disparate physico-chemical nature \cite{Katgert10,Katgertetal09,Becuetal06,Bocquet09,Goyon1,Goyon2,mansard11,mansard13,Ovarlez09,Fielding09,Schall,SchallVanHecke}, such materials share several common characteristics, primarily an anomalous relaxation time to equilibrium  and a non-linear relation between the applied stress and the resulting strain (non-Newtonian rheology) \cite{Sollich1,Sollich2}. These materials show a solid-like behavior at rest and local yielding above  a given applied stress, known as yield stress. Among others, an important question is whether the yield stress can be uniquely determined when the SG materials are not probed in the same way, depending on the different experimental procedures \cite{Moller}. In this Letter, we provide numerical evidence for the former, namely that the yield stress is a property largely independent of the load conditions. The conclusion is supported by extensive numerical simulations based on a lattice Boltzmann (LB) kinetic model of non-ideal binary fluids \cite{CHEM09,EPL10}. The existence of a unique yield stress independent of the forcing mechanism is probed by performing three distinct types of simulations, namely $i)$ Couette flow under time-oscillating load; $ii)$ Couette flow with standard static load; $iii)$ Kolmogorov flow with spatially periodic load. The resulting yield stress fluids can be characterized by the phenomenological Herschel-Bulkley (HB) model. In particular, for the Kolmogorov flow, it is shown that spatial cooperativity exposes a dependence of the global yield stress on the wave-number which leads to a splitting between the local and global yield stress, hence to a distinction between local and global rheology. Such distinction is peculiar to flows with heterogeneous stress, hence it cannot be observed in a Couette flow. Note that the choice of a Kolmogorov flow is instrumental to avoid wall-rheological effects \cite{mansard13}. These effects are relevant for other flows with heterogeneous stresses like, for instance, the Poiseuille flow.

\section{Dynamic rheological model}
\begin{figure*}[t!]
\begin{minipage}{1.4\textwidth}
\includegraphics[width=3.7cm,keepaspectratio]{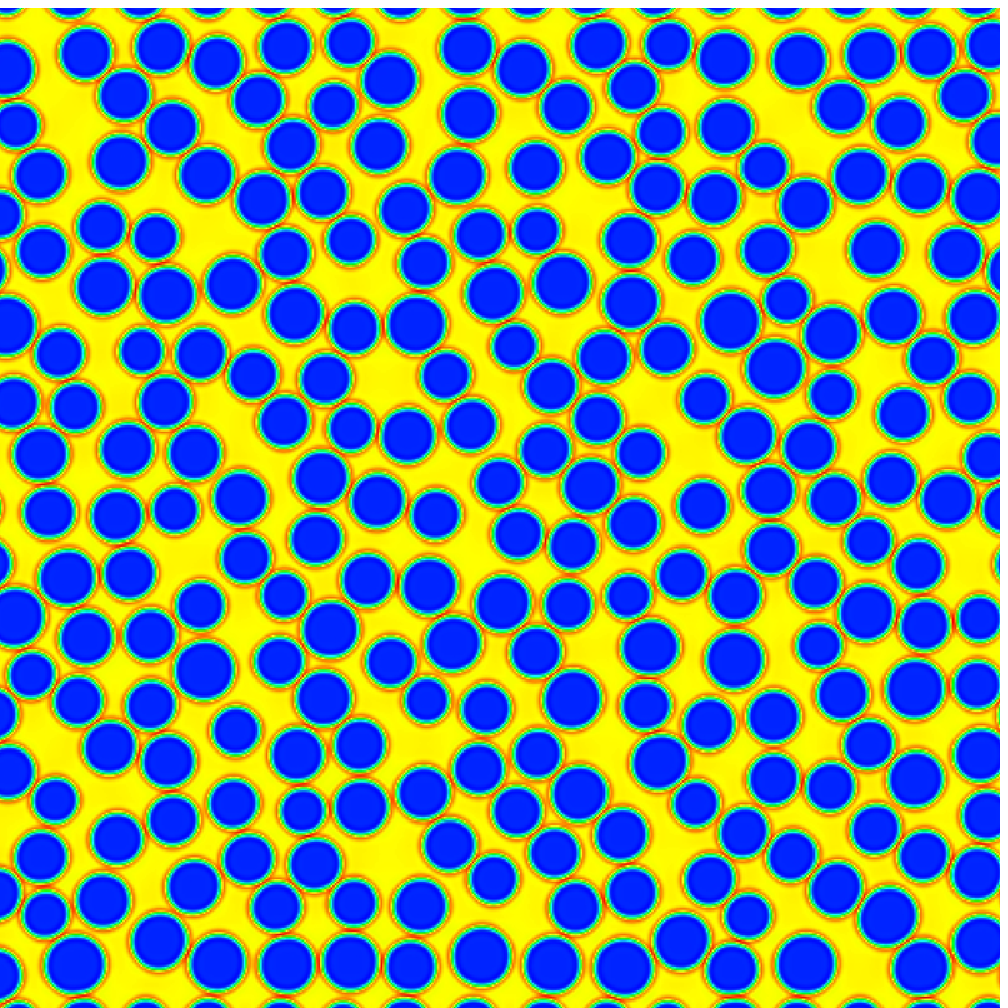}
\includegraphics[width=3.7cm,keepaspectratio]{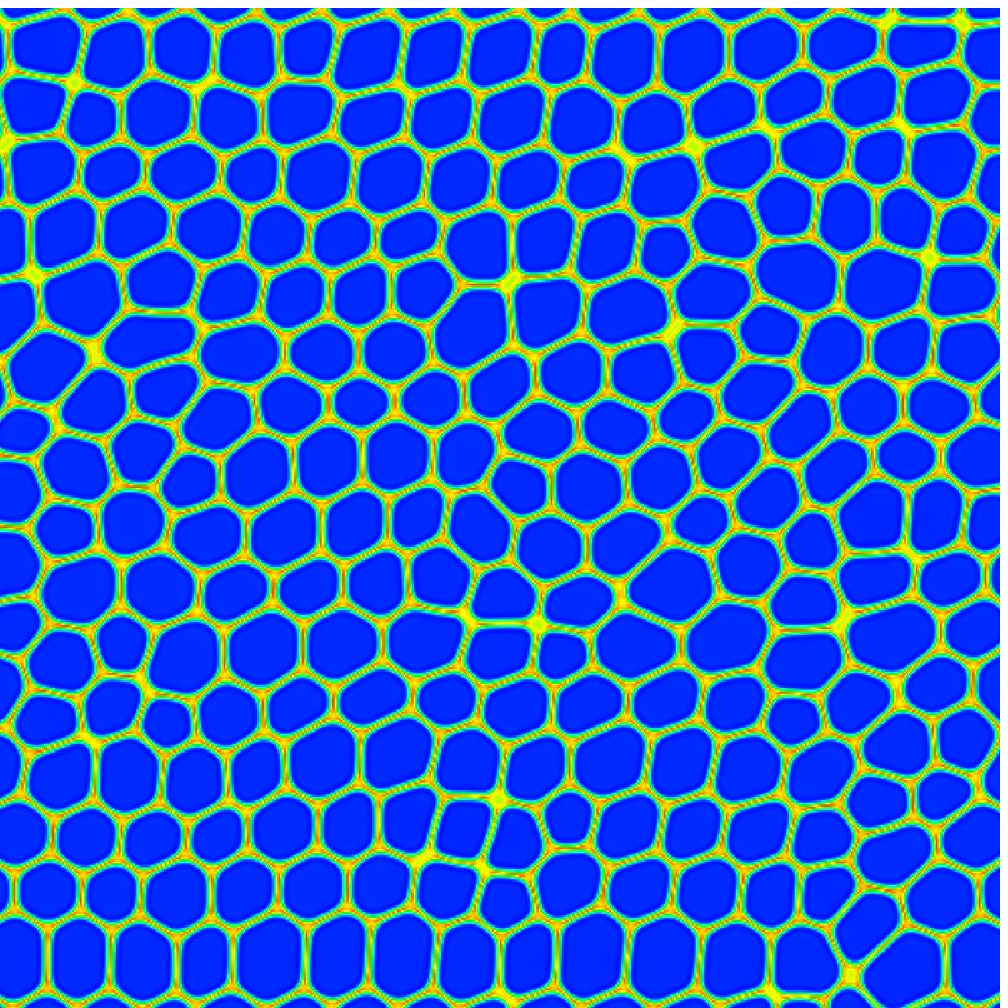}
\includegraphics[width=5.7cm,keepaspectratio]{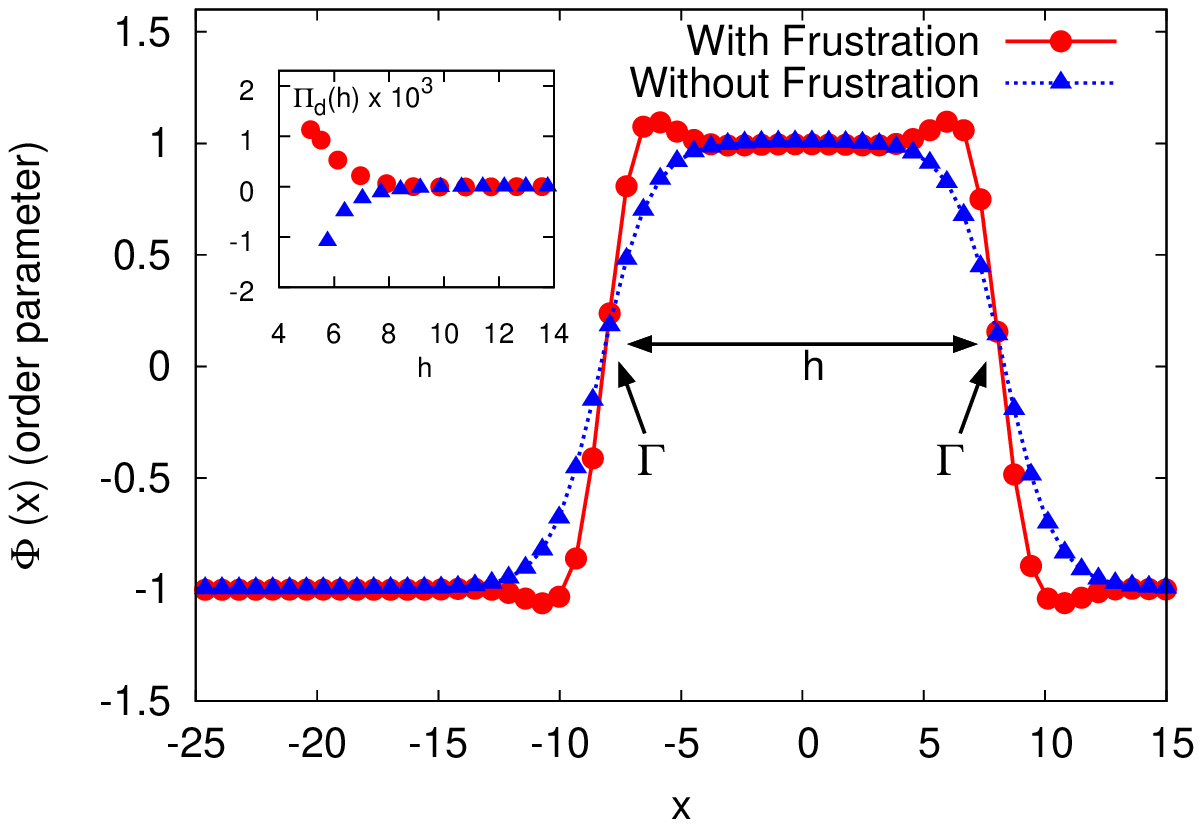}
\includegraphics[width=5.7cm,keepaspectratio]{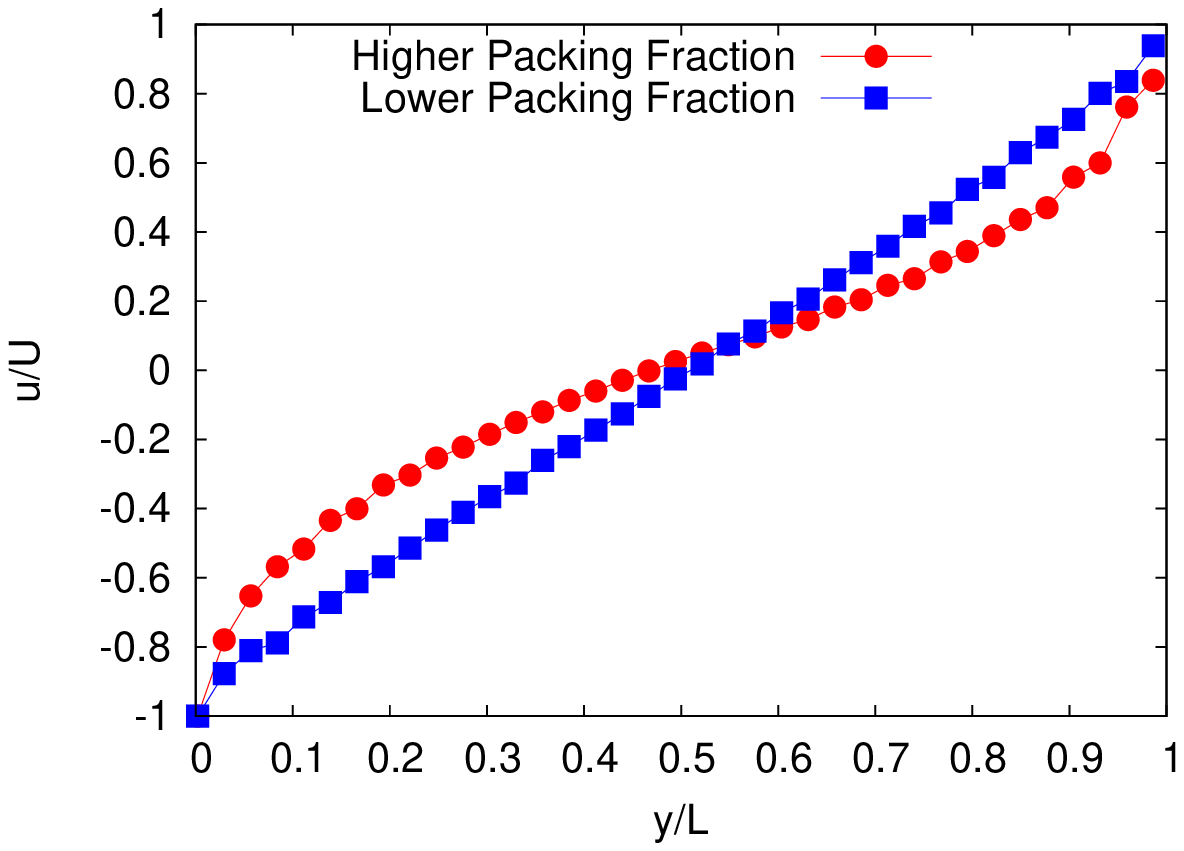}\\
\end{minipage}
\hfill
\begin{minipage}{18cm}
\caption{Panel (a) and (b): droplets simulated with the multicomponent lattice Boltzmann (LB) method. Blue (yellow) colors refer to $A$-rich ($B$-rich) regions. The overall coalescence-stability of these droplets is determined by the stability of the thin liquid films formed between the neighboring drops. Due to the effect of frustration (panel (c)) a positive (stabilizing) disjoining pressure $\Pi_d$ can be achieved for a thin film, the latter represented in terms of the order parameter $\Phi=\rho_A-\rho_B$. Once the droplets are stabilized, different packing fractions and polydispersity (see panels (a) and (b)) of the dispersed phase can be achieved. The effect of the packing fraction is visible in the response of a confined system under the application of a shear $2U/L$ in a channel of with $L$ (panel (d)). The droplet diameter is about $L/20$ and the velocity is averaged in time and in the stream-flow direction. \label{fig1}}
\end{minipage}
\end{figure*}

Here we recall just the essential features of our lattice kinetic model since it has been described in several previous papers \cite{CHEM09,EPL10}. We employ a mesoscopic model of binary fluids, where a suitable frustration mechanism at the interface is introduced by combining a small positive surface tension, promoting highly complex interfaces, with a positive disjoining pressure, inhibiting interface coalescence. The mesoscopic kinetic model considers a binary mixture of fluids $A$ and $B$, each described  by a discrete kinetic Boltzmann distribution function $f_{\zeta i}({\bm r},{\bm c}_i,t)$, measuring the probability of finding a representative particle of fluid $\zeta =A,B$ at position ${\bm r}$ and time $t$, with discrete velocity ${\bm c}_i$, where the index $i$ runs over the nearest and next-to-nearest neighbors of ${\bm r}$ in a regular two-dimensional lattice \cite{BCS,CHEM09}. By definition, the mesoscale particle represents all molecules contained in a unit cell of the lattice. The distribution functions of the two fluids evolve under the effect of free-streaming and local two-body collisions, described, for both fluids ($\zeta=A,B$), by a relation of momentum-relaxation to a local equilibrium ($f_{\zeta i}^{(eq)}$) on a time scale $\tau_{LB}$:
\begin{dmath}\label{LB}
f_{\zeta i}({\bm r}+{\bm c}_i,{\bm c}_i,t+1) -f_{\zeta i}({\bm r},{\bm c}_i,t)  = -\frac{1}{\tau_{LB}} \left(f_{\zeta i}-f_{\zeta i}^{(eq)} \right)({\bm r},{\bm c}_i,t)+F_{\zeta i}({\bm r},{\bm c}_i,t).
\end{dmath}
The equilibrium distribution is given by
\be
f_{\zeta  i}^{(eq)}=w_i \rho_{\zeta} \left[1+\frac{{\bm u} \cdot {\bm c}_i}{c_s^2}+\frac{{\bm u}{\bm u}:({\bm c}_i{\bm c}_i-c_s^2 \id)}{2 c_s^4} \right]
\ee
with $w_i$ a set of weights known a priori through the choice of the quadrature.
The term $F_{\zeta i}({\bm r},{\bm c}_i,t)$ is just the $i$-th projection of  the total internal force which includes a variety of interparticle forces. First, a repulsive ($r$) force with strength parameter ${\cal G}_{AB}$ between the two fluids
\begin{dmath}\label{Phase}
F^{(r)}_\zeta ({\bm r})=-{\cal G}_{AB} \rho_{\zeta }({\bm r}) \sum_{i, \zeta ' \neq \zeta } w_i \rho_{\zeta '}({\bm r}+{\bm c}_i){\bm c}_i
\end{dmath}
is responsible for phase separation \cite{CHEM09} (see triangles in figure \ref{fig1}, panel (c)).  Furthermore, both fluids are also subject to competing interactions whose role is to provide a mechanism for {\it frustration} ($F$) for phase separation \cite{Seul95}. In particular, we model short range (nearest neighbor, NN) self-attraction (controlled by strength parameters ${\cal G}_{AA,1} <0$, ${\cal G}_{BB,1} <0$), and ``long-range'' (next to nearest neighbor, NNN) self-repulsion (regulated by strength parameters ${\cal G}_{AA,2} >0$, ${\cal G}_{BB,2} >0$)
\begin{dmath}\label{NNandNNN}
F^{(F)}_\zeta ({\bm r})=-{\cal G}_{\zeta \zeta ,1} \psi_{\zeta }({\bm r}) \sum_{i \in NN} w_i \psi_{\zeta }({\bm r}+{\bm c}_i){\bm c}_i -{\cal G}_{\zeta \zeta ,2} \psi_{\zeta }({\bm r}) \sum_{i \in NNN} w_i \psi_{\zeta }({\bm r}+{\bm c}_i){\bm c}_i
\end{dmath}
with $\psi_{\zeta }({\bm r})=\psi_{\zeta }[\rho({\bm r})]$ a suitable pseudopotential function \cite{SC1,SbragagliaShan11}.
By a proper tuning of the phase separating interactions (\ref{Phase}) and the competing interactions (\ref{NNandNNN}), the model simultaneously achieves small (positive) surface tension $\Gamma$ and positive disjoining pressure $\Pi_d$ \footnote{The surface tension measured in presence of frustration is smaller with respect to the one without frustration. For the choice of parameters made here it decreases from $\Gamma=0.0846$ (without frustration) to $\Gamma=0.0362$ (with frustration).}. The former promotes the onset of complex density configurations with large surface/volume ratios, whereas the latter frustrates the natural tendency of the interface to minimize its area (length, for the present case of 2d interfaces) via merger events \footnote{We used the same parameters reported in \cite{EPL10} with $\tau_{LB}=1$.}.  The emergence of a positive disjoining pressure $\Pi_d(h)$ is shown in figure \ref{fig1} panel (c), where we consider a thin {\it film} with two non-ideal flat interfaces developing along a given (say $x$) direction, separated by the distance $h$. Following Bergeron \cite{Bergeron}, we write the relation for the corresponding tensions starting from the Gibbs-Duhem relation, $\Gamma_f(h)=2 \Gamma+\int^{\Pi_d(h)}_{\Pi_d(h=\infty)} h \, d \Pi_d$, where $\Gamma_f$ is the overall film tension, whose expression is known in terms of the mismatch between the normal (N) and tangential (T) components of the pressure tensor \cite{Toshev,Derjaguin}, $\Gamma_f=\int_{-\infty}^{+\infty} (P_N-P_T(x)) dx$,  where, in our model, $P_N-P_T(x)=p_s(x)$ can be computed analytically \cite{Shan08}.   Knowledge of the relation $s(h)=\Gamma_f(h)-2 \Gamma$, makes it straightforward to compute the disjoining pressure: a simple differentiation of $s(h)$ permits  to determine the first derivative of the disjoining pressure, $\frac{d s(h)}{d h}=h \frac{d \Pi_d}{d h}$.  This information, supplemented with the boundary condition $\Pi_d(h \rightarrow \infty) = 0$, allows to completely determine the disjoining pressure of the film (see figure \ref{fig1}, inset of panel (c)) \cite{Softmatter}.


\section{Rheology}

We study the rheological properties of the system under three representative conditions: $i)$ Couette flow with oscillating strain at the boundary $U \cos( \omega t)$; $ii)$ Couette flow with steady velocity at the boundaries $\pm U$ and  $iii)$ Kolmogorov flow with external force  $U_0 \eta k^2 \sin(ky)$, $k=2\pi/L$, with periodic boundary conditions. In all cases, the forcing is applied along the flow ($x$) direction.  Upon averaging in $x$, the physics emerging from our lattice kinetic model can be described by the following dynamic rheological model:
\begin{dmath}\label{RHEOM}
\begin{split}
\partial_t (\rho u) & =  \partial_y (\eta S  +  \sigma) + F(y) \\
\partial_t \sigma  & = \rho c^2 S -  \frac{\sigma}{\tau(\sigma)}
\end{split}
\end{dmath}
where $\rho$ is the fluid density, $u$ the mainstream flow speed, $S=\partial_y u$ the shear, $\eta$ the molecular dynamic viscosity, $\sigma$ the non-dissipative component of the stress tensor (ideal and non-ideal pressure plus surface tension), $c^2 \equiv E/\rho$ and $E$ is the elastic modulus.  In (\ref{RHEOM}),  $\tau$ is  a stress-dependent relaxation time, which we assume to define the non linear rheology of the system. The first equation is momentum conservation and the second is a model for the evolution of the stress tensor under the drive of the external force $F(y)$. Models like (\ref{RHEOM}) have been known for long in the literature \cite{Picard02}, where the former equation is usually discarded on account of inertia being totally negligible. For a Couette flow, the total stress tensor $\Pi = \eta S + \sigma$ is constant throughout the system and the {\it effective} viscosity $\eta_{eff}$ is defined as $\eta_{eff} = \eta + \sigma/S$. In most cases, as those presented hereafter, the molecular viscosity $\eta$  is much smaller than the solid contribution and we can estimate $\eta_{eff} \sim \sigma/S  = E \tau \gg \eta$. This ensures a negligible difference between $\Pi$ and $\sigma$. Recently \cite{Goyon1,Goyon2,Bocquet09}, the above model has been extended in such way to make $\tau$ a space-time dependent field (the so-called fluidity $f$, which is the inverse of $\tau$).


\section{Numerical results}

\begin{figure}[h]
\includegraphics[scale=0.65]{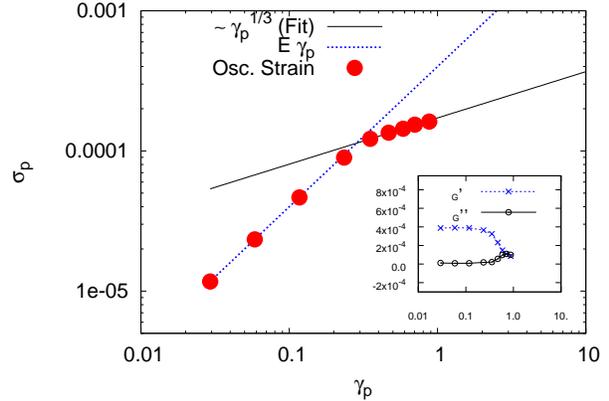}
\caption{The peak stress $\sigma_p$ (filled circles) plotted as a function of the peak strain $\gamma_p$ at $2 \pi / \omega = 1.2 \times 10^{5}$ in lattice Units. In this log-log plot a line with unitary slope is extrapolated through the data at low $\gamma_p$ until it intersects, for the yield stress $\sigma_Y$ and yield strain $\gamma_Y$, with a line of lower slope ($\sim \gamma_p^{1/3}$). In the inset we report the storage modulus, $G^'$,  and loss modulus, $G^{''}$, as a function of peak strain \cite{Larson}.\label{fig2}}
\end{figure}

The computational domain is a square box of size $L \times L$ covered by $N_x \times N_y =1024 \times 1024$ lattice sites. All quantities will be given in LB units. The initial conditions are similar to the one shown in figure \ref{fig1} for the larger packing fraction of the emulsion droplets. The simulations are performed on latest generation Graphics Processing Units (GPU) \cite{GPU}. Close to the yield stress, for both Couette flows and Kolmogorov flow, a reliable statistics can be obtained only by running very time consuming simulations. For Kolmogorov flow at the present resolutions, we perform numerical simulations covering 20 million time steps. In figure \ref{fig2} we show the numerical results obtained for oscillating boundary conditions, with strain $ \gamma(t) = \gamma_p \sin( \omega t)$ and boundary velocity $U(t) = L {\dot \gamma}$. We choose the period $ 2\pi/ \omega$ long enough to attain homogeneous strain in the system for very small $\gamma_p$.   For $L=1024$ and $2 \pi / \omega = 1.2\times10^5$, we found that both the strain $\gamma$ and the stress $\sigma$ are homogeneous in $y$ for very small $\gamma_p$.  We then write $\sigma = \sigma_p \sin( \omega t + \phi)$, where $\sigma_p$ denotes the maximum value of $\sigma$. In figure \ref{fig2}, we show $\sigma_p$ as a function of $\gamma_p$ at  increasing $\gamma_p$. At low values of $\gamma_p$, the elastic relation $\sigma_p = E \gamma_p$ is clearly fulfilled, which allows us to estimate  the elastic modulus as $E=4 \times 10^{-4}$.  At large $\gamma_p$, we found $\sigma_p \sim \gamma_p^{1/3}$.

Similarly to the classical experiments by Mason \cite{mason}, we found a rather sharp change in the two behaviors, which allows the identification of the yield strain $\gamma_Y \sim 0.2 $  and yield stress $\sigma_Y \sim 1.1 \times 10^{-4}$. For $\gamma_p \ge \gamma_Y$, both the strain and the stress are no longer homogeneous in $y$ and, consequently, we compute $\sigma_p$ by averaging $\sigma(y,t)$ in $y$.  Non-homogeneous shear is also observed in steady Couette flow (see also figure \ref{fig1}, panel (d)) for relatively large values of the imposed shear $S=2U/L$, whereas the time average of the stress  $\sigma + \eta \partial_y u$ is constant in space, as it should be.  For the Kolmogorov flow, both $\sigma$ and $S$ are functions of $y$, at any forcing.  At relatively large forcing, we observe a well defined plateau region for small $\sigma$ (see inset of figure \ref{fig6}),   with the existence of a yield-stress $\sigma_Y$.

\begin{figure*}[t!]
\begin{minipage}{1.4\textwidth}
\includegraphics[scale=0.65,keepaspectratio]{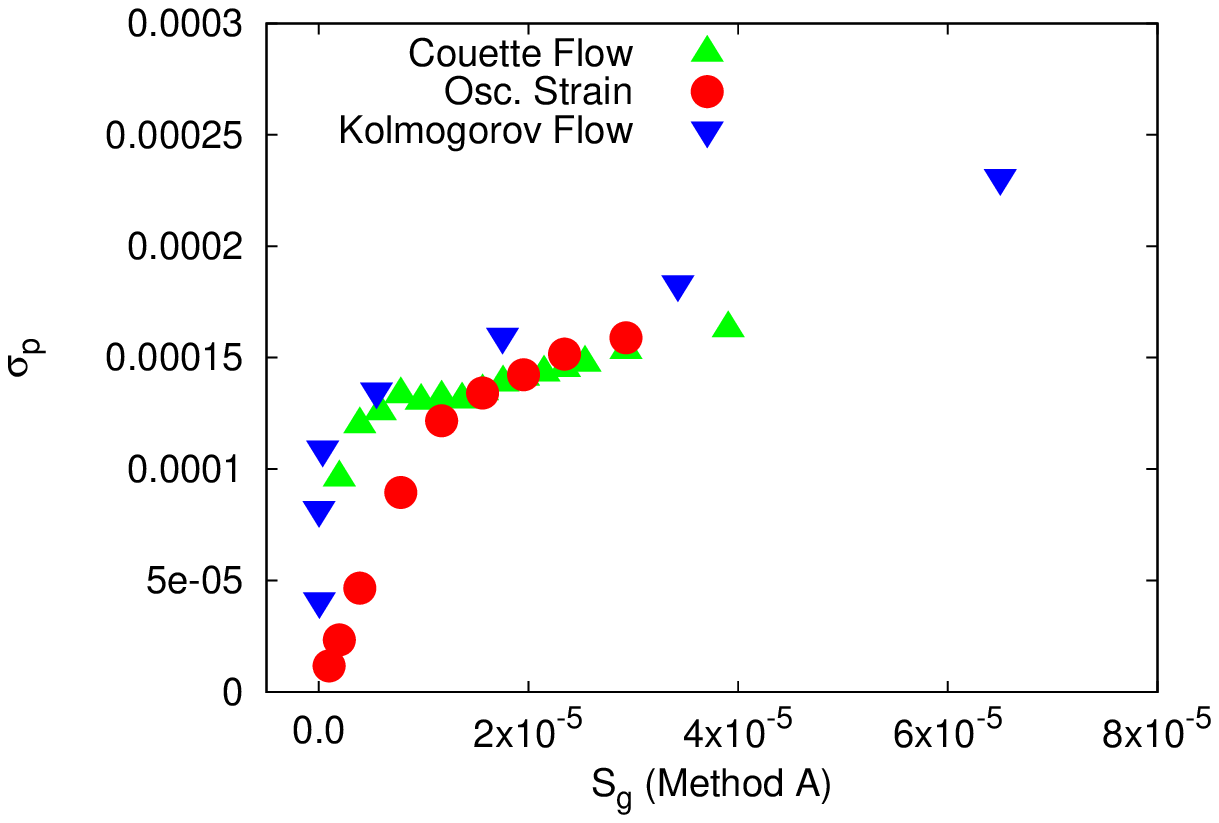}
\includegraphics[scale=0.65,keepaspectratio]{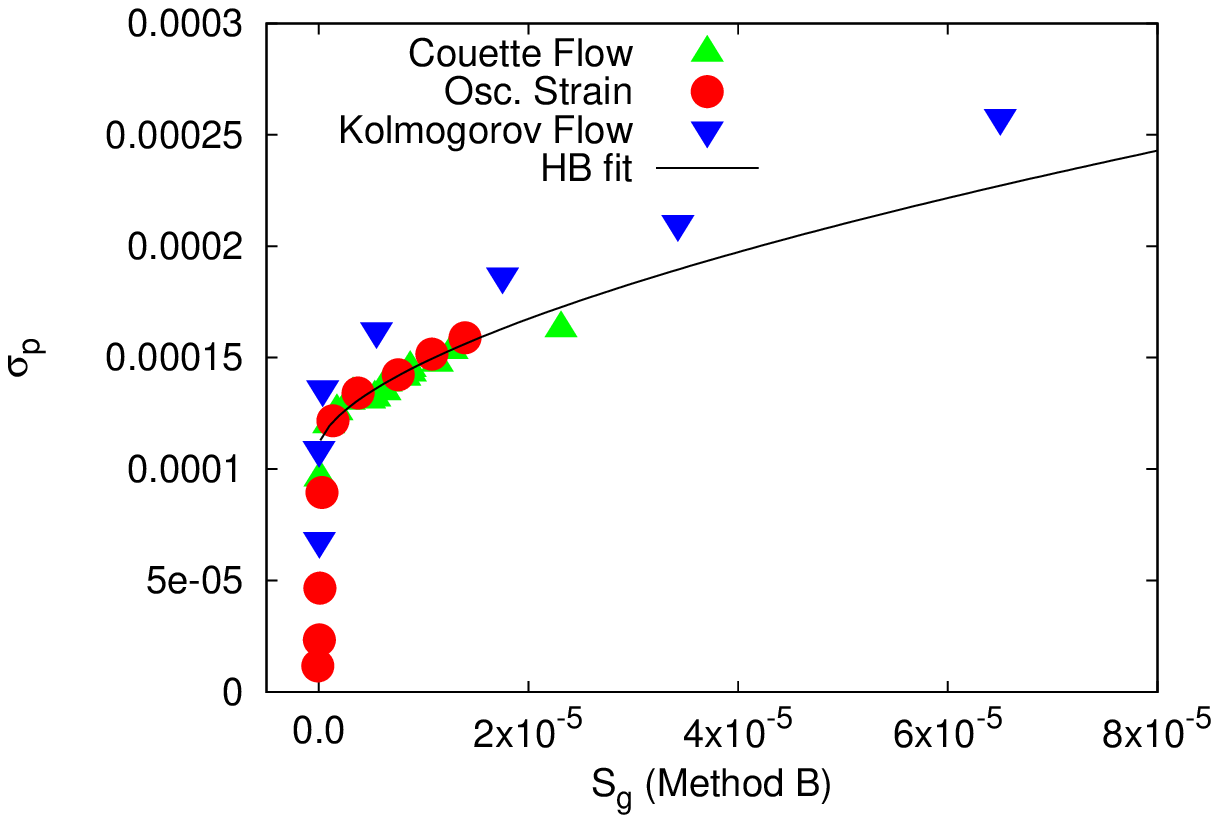}\\
\end{minipage}
\hfill
\begin{minipage}{18cm}
\caption{Global rheological properties of the flow for three different cases: oscillating boundary conditions with strain  $ \gamma(t) = \gamma_p \sin( \omega t)$ and boundary velocity $U(t) = L {\dot \gamma}$ (see also figure \ref{fig2}); Steady Couette flow; Kolmogorov flow with sinusoidal forcing $F(y)=U_0 k^2 \eta \sin (k y)$. Panel (a): the stress $\sigma_p$ is defined as the maximum stress and $S_g$ is the global shear imposed on the system (see text for details).  Panel (b): we report the same data as panel (a) by changing the definition of the global shear. For the oscillating strain and Kolmogorov flow simulations we define the global shear to be $S_{g}=\langle \sigma S \rangle$ /$\sigma_p$, while for the Couette flow we compute $S_{g}$ in the central region where spatial homogeneity is observed (see also figure \ref{fig1}, panel (d)). \label{fig:Reoglobale}}
\end{minipage}
\end{figure*}

We start by investigating the global rheological properties of the flow (figure \ref{fig:Reoglobale}). In all cases, we identify the stress with the maximum stress $\sigma_p$.  For the Couette flow and the oscillating strain we identify the global shear as the external shear imposed to the system $S_{g}=2U/L$ (the peak velocity is considered in the oscillating strain). For the Kolmogorov case, we define $S_g = \langle \partial_y u(y) \partial_y u_0(y) \rangle / (U_0 k)$,  where $u(y)$ is the observed velocity profile, $u_0(y) = U_0 \sin(ky)$, and where brackets stand for space-time averages. This way of defining the global shear is labeled as ``Method A" in the left panel of figure \ref{fig:Reoglobale}. Clearly, the results do not collapse on the same curve, indicating that either the choice of the global variables is not appropriate, or that rheological properties  depend on the forcing mechanism. In order to dig deeper into this matter, we come back to equation (\ref{RHEOM}). Upon multiplying the second equation of (\ref{RHEOM}) by $\sigma$ and integrating in space and time, we obtain $E \langle \sigma S \rangle -  \langle \frac{\sigma^2 }{\tau} \rangle   = 0$. The term $ \langle \sigma S \rangle$ represents the energy transfer (if any) from the kinetic energy of the flow to the elastic stress. In other words, the system starts to flow only when $\langle \sigma S \rangle>0$. Therefore, non trivial rheological properties at the {\it global} level, should appear for a non zero value of $\langle \sigma S \rangle$. This information can be used in the following way: for the oscillating strain and Kolmogorov flow, we define the global shear as $S_{g}=\langle \sigma S \rangle / \sigma_p$, whereas for the Couette flow we compute $S_{g}$ in the central region, where spatial homogeneity is observed. This way of defining the global shear is labeled as ``Method B" in the right panel of figure \ref{fig:Reoglobale}, where the black line  represents the HB fit. Still, a mismatch between the Couette flows and the Kolmogorov flow persists.  A basic difference between such flows must be pointed out: in the former, the shear stress is constant throughout the fluid, so that there is no distinction between local and global rheology. In the latter, however, the stress is a periodic function of the cross-flow coordinate, and consequently  a distinction between local and global rheology opens up. For instance, in the Kolmogorov case, one would expect that the onset of global flow should occur only once the stress exceeds the yield threshold  over an {\it extended} region of the flow. We argue that such region has a linear extension close to the cooperativity length $\xi$ \cite{Goyon1,Goyon2,Geraud13}, recently proposed to characterize the degree of non-locality of stress relaxation phenomena in soft-glassy materials. From this viewpoint, the Kolmogorov flow provides a new opportunity to explore the notion of cooperative length, with no counterpart in the Couette flow. To spot the emergence of the cooperativity phenomenon, we first consider the Couette flow averaged in time and in the stream-flow direction. For a $1d$ case, the fluidity is predicted to obey a non-local equation in the form \cite{Goyon1,Goyon2,Geraud13}:
\be
\xi^2 \frac{d^2 f(y)}{d y^2}+(f_{b}-f(y))=0
\ee
where $f_{b}$ is the bulk fluidity.

\begin{figure}[h]
\begin{center}
\includegraphics[scale=0.65]{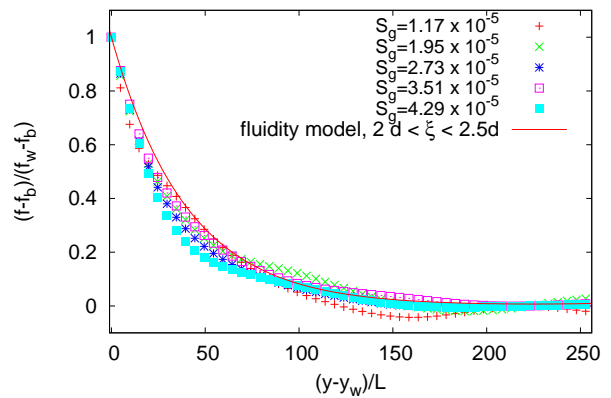}
\caption{Evidence for cooperativity in the stationary Couette flow. The fluidity as a function of the distance from the walls in a Couette Flow simulation.  The fluidity is reported and normalized with respect to the wall (w) and bulk (b) contributions, in order to extract the cooperative length according to equation (\ref{eq:fluidityCF}). Different values of the global shear $S_g$ (as defined in Method B of figure \ref{fig:Reoglobale}) are considered.  \label{fig:fluidity}}
\end{center}
\end{figure}


\begin{figure*}[t!]
\begin{minipage}{1.4\textwidth}
\includegraphics[scale=0.65]{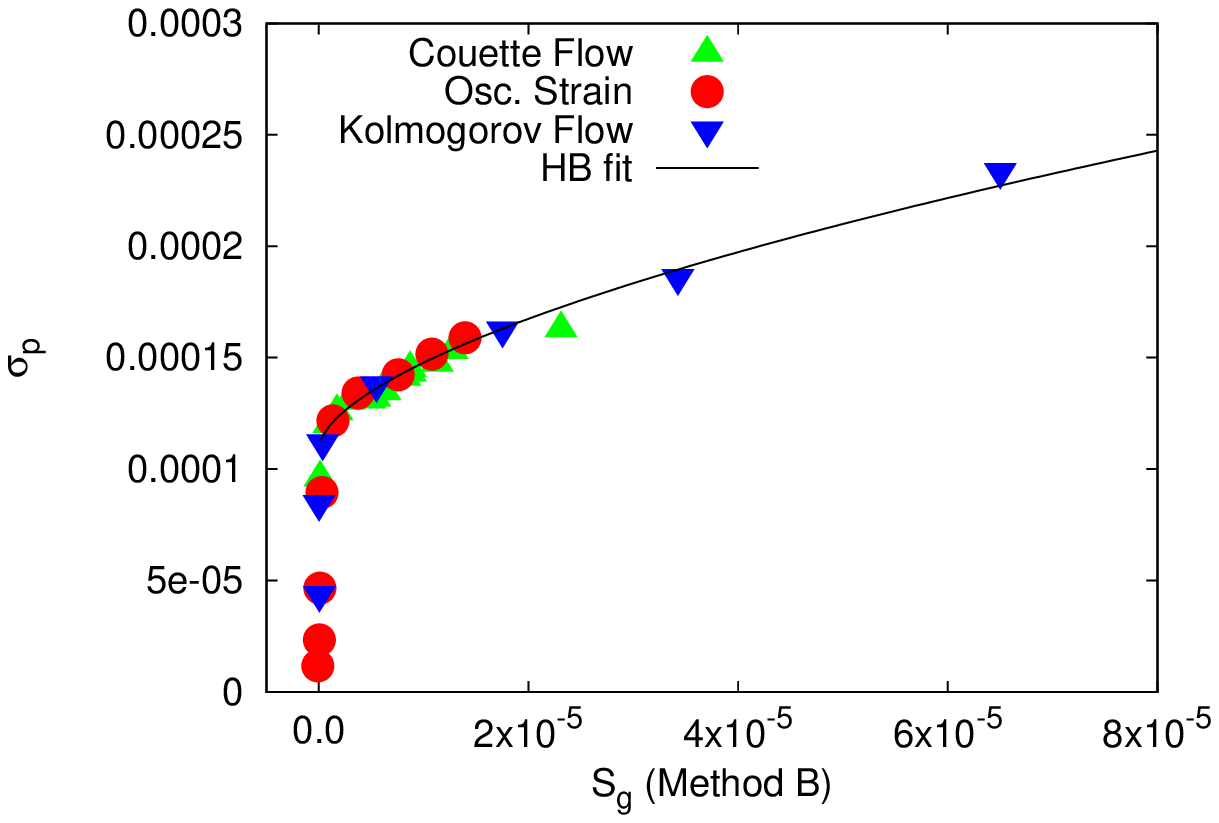}
\includegraphics[scale=0.65]{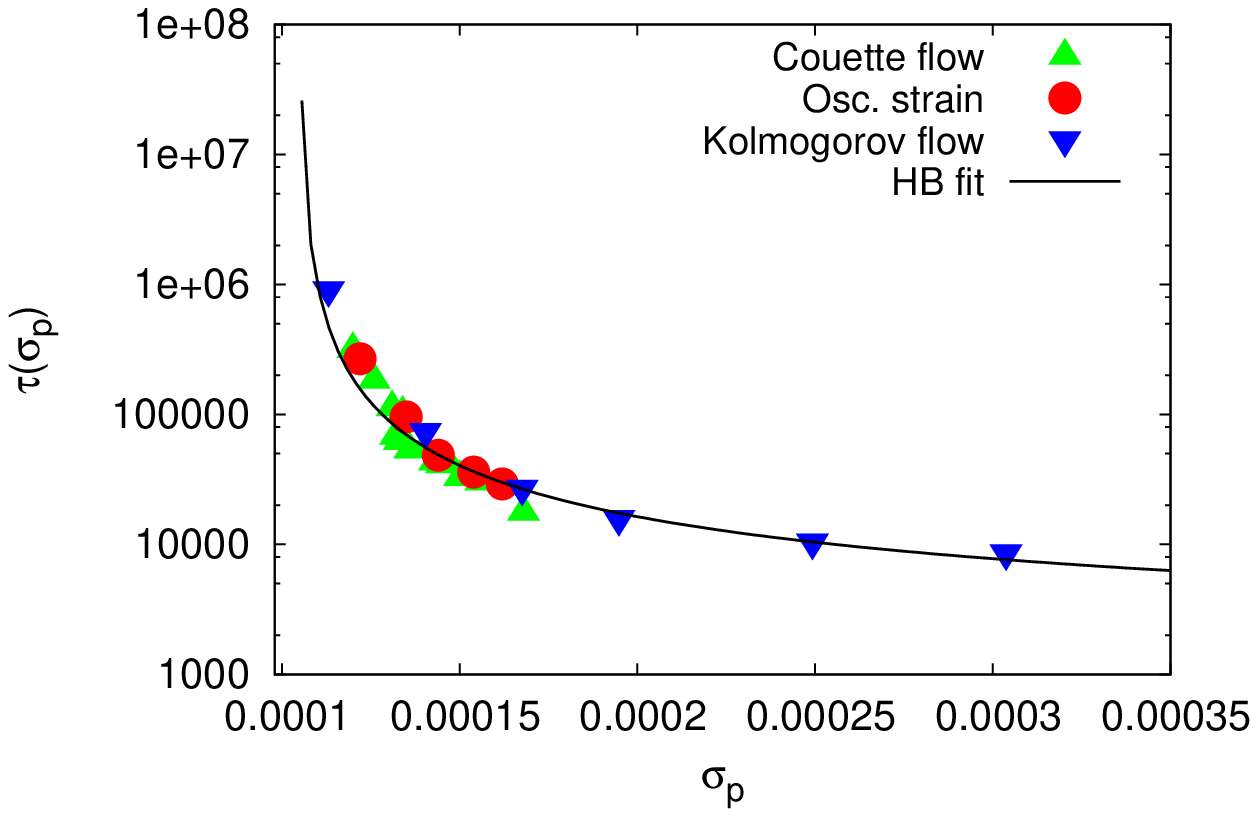}
\end{minipage}
\hfill
\begin{minipage}{18cm}
\caption{Left Panel: Same data reported in the right panel of figure \ref{fig:Reoglobale}, with the stress data for the Kolmogorov flow simulation appropriately rescaled (the rescaling procedure is explained in the text). Right panel: The inverse fluidity $\tau$ as defined by $\tau =  \frac{ \langle \sigma^2 \rangle} {E \langle \sigma S \rangle}$ is computed in the different simulations described in the text: oscillating boundary conditions with strain $ \gamma(t) = \gamma_p \sin( \omega t)$ and boundary velocity $U(t) = L {\dot \gamma}$ (see also figure \ref{fig2}); Steady Couette flow; Kolmogorov flow with sinusoidal forcing $F(y)=U_0 k^2 \eta \sin (k y)$. The different cases provide relations $\tau(\sigma_p)$ collapsing on the same curve consistent with the HB law. \label{fig:correct}}
\end{minipage}
\end{figure*}

The resolution of the fluidity equations requires boundary conditions, i.e. one has to prescribe the fluidity close to the boundaries, $f=f_w$.  Since in the Couette flow the mean shear stress is spatially homogeneous, the shear rate reduces to:
\be\label{eq:fluidityCF}
\dot{\gamma}(y)=\Sigma \left(f_{b}(\Sigma)+(f_w-f_{b}(\Sigma))\frac{\cosh((y-L/2)\xi)}{\cosh(L/2\xi)} \right)
\ee
where $\Sigma = \langle \sigma \rangle$. The fluidity is reported in figure \ref{fig:fluidity}. Starting from the wall region ($y=y_w$), the fluidity field decays towards the bulk value $f_{b}$. To double check the quantitative consistency with equation (\ref{eq:fluidityCF}), we rescaled all the profiles with respect to the wall fluidity $f_w$ and study the quantity $(f(y)-f_{b}(\Sigma))/(f_w-f_{b}(\Sigma))$. Results are consistent with a value of $\xi$ of the order of $2.0 \, d$-$2.5 \, d$, where $d$ is the average  droplet diameter.\\
Let us now go back to the Kolmogorov flow. For a given cross-flow coordinate $y$, the gradient of the stress $\sigma(y)$ (averaged along the stream-flow and time) must counterbalance the forcing because of mechanical equilibrium.  Therefore, we predict $\sigma(y)=U_0 k \eta \cos (k y)$.  To provide a net global motion, one expects the local stress to extend over a region as large as the cooperativity lengthscale $\xi$. Therefore, the peak stress $\sigma^{(eff)}_{Y}$ must be such that there is a region of order $\xi$ above the yield stress $\sigma_Y$
$$
\sigma^{(eff)}_{Y} \cos (k \xi) \approx \sigma_Y.
$$
The net effect is to increase the effective yield stress that is required to support global motion.  A few numbers may help elucidating the picture: for the cooperativity length that we measure (see figure \ref{fig:fluidity}), we obtain $\sigma^{(eff)}_{Y}$ about $20-25 \%$ larger than $\sigma_Y$. Results reported in the right panel of figure \ref{fig:Reoglobale}, with a rescaling of the peak stress for the Kolmogorov flow of such order of magnitude, are reported in the left panel of figure \ref{fig:correct}. From this figure a clear collapse of the numerical data on the same curve is clearly appreciated. Going back to equations (\ref{RHEOM}), by multiplying the second equation by $\sigma$ and by integrating in space and time, we obtain $E \langle \sigma S \rangle - \langle \frac{\sigma^2 }{\tau} \rangle=0$, where brackets stand for space-time averages. We may define a global $\tau$ as $\tau =  \frac{ \langle \sigma^2 \rangle} {E \langle \sigma S \rangle}$ that represents the relaxation time at the global level. In the limit $\sigma \rightarrow \sigma_Y$, we expect $ \tau(\sigma) \rightarrow \infty$. This is actually the case, as shown in the right panel of figure \ref{fig:correct}: the three different cases provide relations $\tau(\sigma_p)$ as a function of $\sigma_p$ which collapse again on the same curve.

In figure \ref{fig6}, we show the stress-shear relation, as obtained from the local values of $\sigma(y)$ and $S(y)$. We note that the yield stress obtained from such local rheology does not show any dramatic departure from the one observed in the Couette case, i.e. $\sigma_Y(y) \approx \sigma_Y$. \footnote{Regarding the relation between this rheological curve and the one obtained in the left panel of figure \ref{fig:correct}, any quantitative statement is left for future investigation}.\\
Finally, we note that the gap between the local ($\sigma_Y$) and global ($\sigma^{(eff)}_{Y}$) yield stress is expected to lead to intermittent flow, whenever the actual stress falls between the local and global yield threshold. This is exactly the case, as evidenced by figure \ref{fig:intermittenza}, where we report the Fourier component $U(k,t)$ of the stream-flow averaged velocity field, normalized  in such a way that $U(k,t)=U_0$ in the case of a purely Newtonian fluid. Simulations are performed for a Kolmogorov flow with $U_0=0.1$ and a peak stress $\sigma_p$ such that $\sigma_Y < \sigma_p < \sigma^{(eff)}_{Y}$.  To provide some reference number, we have taken the shear corresponding to $\sigma_p$ from the HB relation as  obtained in the Couette flow simulation, and extracted the representative velocity for the Couette flow (its wall velocity) corresponding to such stress.

\begin{figure}[h!]
\includegraphics[scale=0.65]{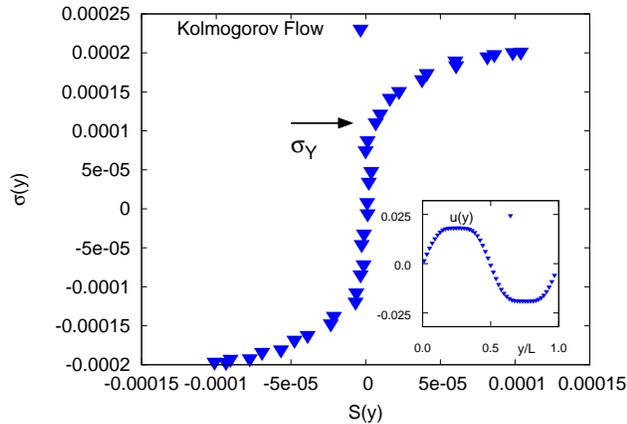}
\caption{Stress-shear relation for a Kolmogorov flow simulation. The local values are extracted from the numerical simulations by averaging in time and along the stream-flow direction (see inset). The yield stress $\sigma_Y$ obtained for the Couette flow and the Oscillating strain simulations is indicated with the arrow. \label{fig6}}
\end{figure}

\begin{figure}[h!]
\includegraphics[scale=0.65]{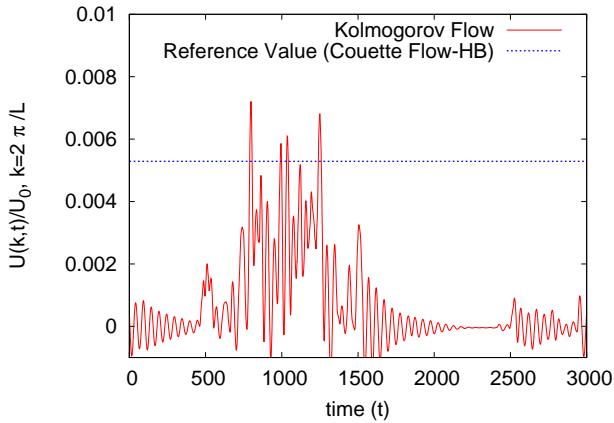}
\caption{We report the Fourier component $U(k,t)$ of the stream-flow averaged velocity (see text for details). Simulations are performed for a Kolmogorov with a peak stress $\sigma_p$ such that $\sigma_Y < \sigma_p < \sigma^{(eff)}_{Y}$. \label{fig:intermittenza}}
\end{figure}



\section{Summary and outlook}
We have shown that LB simulations provide an extremely versatile tool to investigate non linear rheology in different flow configurations including Couette flow, time-oscillating Strain and Kolmogorov flow.  In all cases, the global rheology is described by the HB relation, with the yield stress largely independent of the loading scenario. To this purpose, rescaling with an effective yields stress (based on the notion of cooperativity \cite{Goyon1,Goyon2,Geraud13}) proves instrumental in recovering a universal shear-stress relation for both stress-homogeneous (Couette) and inhomogeneous (Kolmogorov) flows. A systematic investigation of the role of non-local effects, in the spirit of the work described by \cite{Goyon1,Goyon2,Bocquet09}, is definitely warranted for future investigations. In particular, preliminary results suggest a direct link between the cooperative length $\xi$ and the characteristic size of plastic events. M.S. \& R.B.  kindly acknowledge funding from the European Research Council under the European Community's Seventh Framework Programme (FP7/2007-2013)/ERC Grant Agreement no[279004]. We thank the PRACE Distributed European Computing Initiative (DECI-8) for providing us access to CINECA computing resources.

\end{document}